\documentclass[11pt]{article}

\makeatletter
\@addtoreset{equation}{section}

\makeatother

\def\Tr{\mathop{\rm Tr}\nolimits}
\def\bar#1{\overline{#1}}
\def\Eq#1{Eq.\ (\ref{#1})}
\newcommand{\Eqs}[2]{Eqs.\ (\ref{#1}) and (\ref{#2})}
\newcommand{\Eqss}[3]{Eqs.\ (\ref{#1}), (\ref{#2}) and (\ref{#3})}

\def\bm#1{\mbox{\boldmath $#1$}}
\def\diag#1{\mathop{\rm diag} \Big( #1 \Big)}
\def\ddiag#1{\mathop{\rm diag}^{\longleftarrow} \Big( #1 \Big)}

\font\bb=bbmss10 scaled 1200
\def\umat{\mbox{\bb 1}}

\title{\centerline{\normalsize June 2003 \hfill SINP/TNP/03-16}\bigskip
\bf Generalized Fierz identities}

\author{
\bf Jos\'e F. Nieves\\
Laboratory of Theoretical Physics\\ 
Department of Physics, P.O. Box 23343\\
University of Puerto Rico, R\'{\i}o Piedras,
Puerto Rico 00931-3343
\and
\bf Palash B. Pal\\
Saha Institute of Nuclear Physics\\ 
1/AF Bidhan-Nagar, Calcutta 700064, India
}

\date{}

\begin{document}
\maketitle

\begin{abstract}
Low energy weak interactions calculations with fermions frequently
involve a superposition of quartic products of Dirac spinors, in which
the order of the spinors is not the same in all the contributing
terms.  A common trick that is used to bring them to a uniform
ordering is the Fierz transformation.  We show that the standard Fierz
rearrangement formula quoted in textbooks is one element of a class of
transformations of a quartic product amplitude, under which the
spinors are rearranged with different orderings and, in the general
case, some or all of the spinors are transformed to their
Lorentz-invariant complex conjugate form.  We give a pedagogical
derivation of the explicit forms of all such transformation matrices.
In addition to the usual Lorentz scalar quartic products, we consider
pseudoscalar ones as well.  Such manipulations and formulas are
useful, in particular, when some of the fermions involved are Majorana
particles.

\end{abstract}

\section{Introduction}
\subsection{What are the generalized Fierz identities}
Dirac bilinears are matrix elements of the Dirac $\gamma$ matrices and
products of the $\gamma$ matrices, taken between two spinors.  Such
bilinears appear in any calculation involving fermions and usually the
spinors are the plane wave solutions of the Dirac equation. In
general, the spinors correspond to different momenta and/or different
particles, and therefore are different.  Typical calculations involve
the product of two such bilinears and, since they contain four
spinors, we will call them quadrilinears for the sake of brevity.
Although the order of the two bilinears in the quadrilinear product is
not important, the order of appearance of the individual spinors is.

The Fierz identities \cite{Fierz} are relations between the
quadrilinears, written with two different orders of the spinors.  It
is easy to count how many such different orderings are possible.
Suppose we start with a quadrilinear in the form
\begin{eqnarray}
\Big[ \bar w_1 M w_2 \Big] \Big[ \bar w_3 M' w_4 \Big] \,,
\end{eqnarray}
where $M$ and $M'$ are numerical matrices and the $w_n$ denote four
Dirac spinors. Each $w_n$ can be either a positive energy $u$-spinor
or a negative energy $v$-spinor and the numerical subscript $n$ labels
the combination of momentum (including the mass) and spin that defines
the spinor.  If we want to interchange the places of these spinors,
the same product will be expressed by some different matrices between
the spinors.  Obviously, there can be $4!=24$ different orderings.
However, since the ordering of the bilinears is irrelevant as we have
said, half as many of these orderings would give rise to new
quadrilinears.  There can therefore be 12 different orderings.  One of
these will be the same as the order that we start with, so there are
11 alternative orders left.  The usual Fierz transformation refers to
only one of them, in which $w_2$ and $w_4$, or equivalently $w_1$ and
$w_3$, are interchanged.  In addition, the standard identities involve
combinations of products of bilinears such that the quadrilinears
involved are Lorentz scalars.  In this article, we give explicit
formulas for all 11 orderings, and we also consider additional
quadrilinear products besides the usual ones. To state this more
precisely we introduce some notation.

\subsection{Notation for Dirac matrices}
As is well known, starting with the standard $4\times 4$ Dirac
matrices $\gamma^\mu$ and taking suitable products of them, a set of
16 matrices can be constructed which, among other properties, span the
16 dimensional space of all the $4\times 4$ matrices.  In the notation
that we will use in the subsequent material, a conventional choice of
the 16 matrices is
\begin{eqnarray}
\label{defGam}
\begin{array}{r@{\mbox{ through }}l@{\null\equiv\null}l}
\multicolumn{2}{r@{\null\equiv\null}}{\Gamma_S^1} &  \umat \,, \\ 
\Gamma_V^{1} & \Gamma_V^{4} & \gamma^\mu \,, \\ 
\Gamma_T^1 & \Gamma_T^{6} &  
\sigma^{\mu\nu}  \,,  \quad (\mu<\nu) \\  
\Gamma_A^{1} & \Gamma_A^{4} &  i\gamma^\mu \gamma_5 \,, \\ 
\multicolumn{2}{r@{\null\equiv\null}}{\Gamma_P^{1}} &  \gamma_5 \,,
\end{array}
\end{eqnarray}
where
\begin{eqnarray}
\sigma_{\mu\nu} &=& {i\over 2} \Big[ \gamma^\mu, \gamma^\nu \Big] \,,\\*
\gamma_5 &=& i \gamma^0 \gamma^1 \gamma^2 \gamma^3 \,.
\end{eqnarray}
Further, for definiteness, we adopt the convention that
\begin{eqnarray}
\label{anticomm}
\Big\{ \gamma^\mu, \gamma^\nu \Big\} = 2g^{\mu\nu} \umat \,,
\end{eqnarray}
where the curly braces denote the anticommutator, $\umat$ is the unit
matrix, and the metric tensor is given by
\begin{eqnarray}
g_{\mu\nu} = \mathop{\rm diag} (+1,-1,-1,-1) \,.
\end{eqnarray}
Then it is easily verified that 
\begin{eqnarray}
\Tr \left( \Gamma_I^r \Gamma_{Jq} \right) = 4\delta_{IJ}\delta^r_q \,,
\label{trace}
\end{eqnarray}
where the matrices $\Gamma_{Ir}$ are defined by lowering the Lorentz
indices in Eq (\ref{defGam}).  One more convention we adopt is the
following. Whenever the Lorentz indices (e.g. $r,q$ in the above
formula) appear repeated in a given formula, the usual summation
convention over repeated indices is assumed. On the other hand, we do
not use that convention for the indices $I,J$, which take the
\emph{values} $S,V,T,A,P$. Any summation over these indices will
always be indicated explicitly.

A useful set of relations are the generalizations of the formulas such as 
\begin{eqnarray}
\gamma^\mu\gamma^\alpha\gamma_\mu & = & -2\gamma^\alpha \nonumber\\
\gamma^\mu\sigma^{\alpha\beta}\gamma_\mu & = & 0 \nonumber\\
\sigma^{\mu\nu}\sigma_{\mu\nu} & = & 12 \,,
\end{eqnarray}
and similar ones which follow very simply by repeated application 
of \Eq{anticomm}. Thus, writing
\begin{equation}
\label{Gammacontractions}
\Gamma_I^q \Gamma_J^s \Gamma_{Iq} = f_{IJ}\Gamma_J^s \,,
\end{equation}
they can be summarized in the form
\begin{eqnarray}
\label{f}
\bm f = \left(\begin{array}{rrrrr}
1 &  1 & 1 & 1 & 1 \\
4 & -2 & 0 & 2 & -4 \\
6 &  0 & -2 &  0 & 6 \\
4 & 2 & 0 & -2 & -4 \\
 1 & -1 & 1 &  -1 & 1
\end{array}\right) \,.
\end{eqnarray}
Another useful relation that we quote for future purposes is the
identity
\begin{equation}
\label{Tdual=TP}
\frac{i}{2}\epsilon^{\mu\nu\alpha\beta}\sigma_{\alpha\beta} = 
\sigma^{\mu\nu}\gamma_5 \,,
\end{equation}
with the convention that
\begin{equation}
\epsilon^{0123} = +1 \,.
\end{equation}

\subsection{Dirac quadrilinears and Fierz identities}
In correspondence with the 16 Dirac matrices $\Gamma^r_I$ we then
construct the following bilinears of two spinors $w_1$ and $w_2$,
\begin{equation}
\label{bilinears}
e^r_I(12) = \bar w_1\Gamma^r_I w_2 \,.
\end{equation}
The simplest quadrilinears are products of the form $e_I^r(12)
e_{Ir}(34)$.  In anticipation of obtaining the final formulas in a
convenient form we define
\begin{eqnarray}
e_I(1234) = n_I^2 e_I^r(12) e_{Ir}(34) \,,
\label{e}
\end{eqnarray}
with the numerical coefficients $n_I$ chosen to be
\begin{eqnarray}
n_I = \cases{1 & if $I=S,V,P$, \cr
-i & if $I=A$,\cr 
\surd 2 & if $I=T$,}
\label{n_I}
\end{eqnarray}
so that the quadrilinears are given in terms of the Dirac matrices
simply as
\begin{eqnarray}
e_S(1234) & = & 
\Big[ \bar w_1 w_2 \Big] \Big[ \bar w_3 w_4 \Big]\,,\nonumber\\
e_V(1234) & = &
\Big[ \bar w_1 \gamma^\mu w_2 \Big]
\Big[ \bar w_3 \gamma_\mu w_4 \Big]\,,  \nonumber\\ 
e_T(1234) &=& \Big[ \bar w_1 \sigma_{\mu\nu}
w_2 \Big] \Big[ \bar w_3 \sigma^{\mu\nu} w_4 \Big] \,,\nonumber\\
e_A(1234) &=& \Big[ \bar w_1 \gamma^\mu \gamma_5 w_2 \Big]
\Big[ \bar w_3 \gamma_\mu \gamma_5 w_4 \Big]\,,  \nonumber\\
e_P(1234) & = &
\Big[ \bar w_1 \gamma_5 w_2 \Big]\Big[ \bar w_3 \gamma_5 w_4 \Big]\,.
\end{eqnarray}
In this notation, the standard Fierz transformation
gives the relation between the $e_I(1234)$ and the $e_J(1432)$.

As already mentioned, in this article, we generalize the standard
Fierz transformation in two ways. Firstly, we give the explicit
formulas for the relation between the $e_I(1234)$ and the $e_I(abcd)$
for all the 11 possible final orderings. Secondly, we consider more
general quadrilinear products of the form
\begin{equation}
\label{genquad1}
e_I^r(12) e_{Jr}(34) \,,
\end{equation}
with $I \not= J$, but for those pairs of $I,J$ for
which the Lorentz indices match and are contracted. Thus, the
possibilities are $(I,J) = (V,A)$ and $(I,J) = (S,P)$.
In addition we consider another possible contraction
\begin{equation}
\label{genquad2}
e^r_T(12) e_{\tilde T r}(34)\,,
\end{equation}
where $e^r_{\tilde T}(ab)$ is defined as $e^r_{T}(ab)$ but with
the gamma matrices
\begin{eqnarray}
\label{defTdual}
\begin{array}{r@{\mbox{ through }}l@{\null\equiv\null}l}
\Gamma_{\tilde T}^1 & \Gamma_{\tilde T}^{6} &  
\frac{i}{2}\epsilon^{\mu\nu\alpha\beta}\sigma_{\alpha\beta}\,,  
\quad (\mu<\nu) 
\end{array} \,.
\end{eqnarray}
%

\section{Fierz identities for scalar combinations}
\label{s:scalar}
\subsection{Derivation of the usual Fierz identity}
The usual Fierz identities are relations of the type
\begin{equation}
\label{fierzstandard}
\label{1234}
e_I(1234) = \sum_J F_{IJ}e_J(1432) \,,
\end{equation}
where the $F_{IJ}$ are numerical coefficients. The fact that they must
exist follows from two properties of the generalized gamma
matrices. The first is the closure relation
\begin{equation}
\label{closure}
\frac{1}{4}\sum_J \left(\Gamma_J^r\right)_{ad}
\left(\Gamma_{Jr}\right)_{cb} = 
\delta_{ab}\delta_{cd} \,,
\end{equation}
which is easily proven by noticing that any $4\times 4$ matrix $M$ 
must be expandable in terms of the gamma matrices in the form
\begin{equation}
M = \sum_I m_{Ir} \Gamma_I^r \,,
\end{equation}
where the orthogonality relation in \Eq{trace} implies
that the coefficients are given by
\begin{equation}
m_{Ir} = \frac{1}{4}\Tr\Gamma_{Ir} M \,.
\end{equation}
Therefore
\begin{equation}
M = \frac{1}{4}\sum_I \Gamma_I^r \left(\Tr\Gamma_{Ir} M\right)\,,
\end{equation}
and since this holds for any matrix $M$, the closure relation follows.

The second property is that for any two gamma matrices $\Gamma_I^q$
and $\Gamma_J^r$, there is a third matrix $\Gamma_K^s$ such that
\begin{equation}
\label{productrule}
\Gamma_I^q \Gamma_J^r = \Gamma_K^s \times \mbox{number}\,.
\end{equation}
In general, for any other choice of basis, the right-hand side in
\Eq{productrule} would involve a linear combination of all the basis
matrices. The distinguishing feature of \Eq{productrule} is that only
one member of the basis appears on the right-hand side.

Multiplying \Eq{closure} by 
\begin{equation}
\left(\Gamma_I^q\right)_{a'a} \left(\Gamma_{Iq}\right)_{c'c}
\end{equation}
gives
\begin{equation}
\frac{1}{4}\sum_J \left(\Gamma_I^q\Gamma_J^r\right)_{ad} 
\left(\Gamma_{Iq}\Gamma_{Jr}\right)_{cb} = 
\left(\Gamma_I^q\right)_{ab}\left(\Gamma_{Iq}\right)_{cd} \,,
\end{equation}
and using \Eq{productrule} it implies that there is a relation of
the form
\begin{equation}
\label{fierzstandard2}
\left(\Gamma_I^q\right)_{ab}\left(\Gamma_{Iq}\right)_{cd} = 
\sum_K C_{IK}\left(\Gamma_K^r\right)_{ad} 
\left(\Gamma_{Kr}\right)_{cb} \,,
\end{equation}
which is in fact the Fierz identity in disguise.  Notice that we have
used the fact that, since the Lorentz indices in the quantity on the
left-hand side are contracted such that the quantity is a scalar, the
same must hold on the right-hand side.  The coefficients $C_{IJ}$ can
be found by multiplying \Eq{fierzstandard2} by
\begin{equation}
\left(\Gamma_{Js}\right)_{da} \left(\Gamma_J^s\right)_{bc}
\end{equation}
and using \Eq{trace} once more, which yields
\begin{equation}
C_{IJ} = \frac{1}{16N_J}\Tr \Gamma_{Js} \Gamma_I^q
\Gamma_J^s\Gamma_{Iq} \,, 
\end{equation}
where $N_J$ is the number of gamma matrices in each group $J$, i.e.,
$N_S = N_P = 1$, $N_T = 6$ and $N_V = N_A = 4$. The traces in \Eq{CIJ}
can be evaluated simply by using \Eqs{Gammacontractions}{trace}, and
thus we obtain
\begin{equation}
\label{CIJ}
C_{IJ} = \frac{1}{4}f_{IJ}\,,
\end{equation}
with the $f_{IJ}$ given in \Eq{f}.  Remembering the non-trivial
factors of $n_I$ in the definitions of $e_T$ and $e_A$ given in
\Eq{n_I}, the relation given in \Eq{fierzstandard2} implies
\Eq{fierzstandard}, with the coefficients being
\begin{eqnarray}
F_{IJ} = {n_I^2 \over n_J^2} \; C_{IJ} \,.
\end{eqnarray}
In other words, the matrix $\bm F$ is given by
\begin{eqnarray}
\bm F = 
\frac14 \left(\begin{array}{rrrrr}
 1 &  1 & \frac12 & -1 & 1 \\
 4 & -2 & 0       & -2 & -4 \\
12 &  0 & -2      &  0 & 12 \\
-4 & -2 & 0       & -2 & 4 \\
 1 & -1 & \frac12 &  1 & 1
\end{array}\right) \,,
\label{F}
\end{eqnarray}
which is the usual Fierz transformation matrix.  We will use the
matrix notation
\begin{eqnarray}
\label{fierzstandardshort}
\bm e(1234) = \bm F \bm e(1432)
\end{eqnarray}
as a short hand for \Eq{fierzstandard}.

\subsection{Other orderings}
In the relation appearing in \Eq{1234}, the nature of the exchanged
spinors remains the same.  In other words, a $u$-spinor remains a
$u$-spinor, and the same applies for $v$-spinors.  In fact, if we
restrict the transformations to those for which this property holds,
there are no other possible rearrangements of the spinors.  Here we
contemplate other transformations which do not have that property.

For illustration, we consider the particularly simple example where we
want to interchange the positions of the spinors in the first and
second place.  Clearly this is not possible unless we change a
$u$-spinor to a $v$-spinor and vice versa. Therefore, the appropriate
quantities to consider are
\begin{eqnarray}
e_I(2^c 1^c 34) = 
\Big[ \bar{w^c}_2 \Gamma^r_I w^c_1 \Big] \Big[ \bar w_3 \Gamma_{Ir}
w_4 \Big] \,,
\end{eqnarray}
were we have introduced the notation $w^c$.  If $w$ is a $u$-spinor,
$w^c$ denotes the corresponding $v$-spinor, and vice versa.  In either
case, $w$ and $w^c$ are related by
\begin{eqnarray}
w^c = \gamma_0 C w^* \,,
\label{what}
\end{eqnarray}
or equivalently
\begin{eqnarray}
\bar{w^c} = w^\top C^{-1} \,,
\label{whatbar}
\end{eqnarray}
where $C$ is the conjugation matrix which satisfies the relations
\begin{equation}
\label{Ctranspose}
C^\top = \null -C \,,
\end{equation}
and
\begin{eqnarray}
\label{defC}
C^{-1} \Gamma_I^r C = \eta_I \Gamma_I^{r\top}  \,,
\end{eqnarray}
with the values $\eta_I$ are summarized in the following table:
\begin{eqnarray}
\begin{tabular}{c||rrrrr}
$I$ & S & $V$ & $T$ & $A$ & $P$ \\ 
\hline
$\eta_I$ & $+1$ & $-1$ & $-1$ & $+1$ & $+1$ \\
\end{tabular}
\end{eqnarray}
Therefore, using the fact that the $w$'s are numerical spinors,
by simple matrix algebra we can write
\begin{eqnarray}
\bar{w^c_2} M w^c_1 
= \null - \bar w_1 C M^\top C^{-1} w_2\,,
\label{transpose}
\end{eqnarray}
for any $4\times4$ matrix $M$. In particular, using
\Eqs{Ctranspose}{defC},
\begin{eqnarray}
\label{transposeGamma}
\bar w_1 \Gamma^r_I w_2 = \null -\eta_I \, \bar{w^c_2} \Gamma^r_I
w^c_1 \,. 
\end{eqnarray}
If the $w$'s were to denote anticommuting fields, the relations
analogous to \Eqs{transpose}{transposeGamma} would not have the
overall minus sign on the right-hand side.

It follows from \Eq{transposeGamma} that the Fierz transformation
matrix $\bm S$ defined by the relation
\begin{eqnarray}
e_I (1234) = \sum_J S_{IJ} e_J(2^c 1^c 34) \,,
\label{defS}
\end{eqnarray}
is simply
\begin{eqnarray}
\bm S = \diag {-1,+1,+1,-1,-1} \,,
\label{S}
\end{eqnarray}
and it is clear that the same matrix appears in the relation
\begin{eqnarray}
e_I (1234) = \sum_J S_{IJ} e_J(124^c 3^c) \,.
\label{S34}
\end{eqnarray}

Combining the standard Fierz transformations with \Eqs{defS}{S34}, the
transformations for the other permutations are simple to obtain.  For
example, consider the interchange $2\leftrightarrow3$.  Using \Eq{F}
followed by \Eq{defS}, we obtain
\begin{eqnarray}
\bm e(1234) = \bm F \bm e(1432) 
= \bm F \bm S \bm e(142^c 3^c)
= \bm F \bm S \bm F \bm e(13^c 2^c 4) \,.
\end{eqnarray}
Alternatively, we can write 
\begin{equation}
\bm e(1234) = \bm {Se}(2^c1^c 3 4) = \bm{SFe}(2^c 4 3 1^c) 
= \bm{SFSe}(13^c2^c 4) \,,
\end{equation}
which shows that
\begin{eqnarray}
\bm{SFS} = \bm{FSF} \,.
\label{SFS=FSF}
\end{eqnarray}
Proceeding in similar fashion, the Fierz matrices for all the
interchanges are obtained.  We denote the initial ordering always by
(1234), and any of the final re-orderings, such as those we have
considered explicitly, by $(abcd)$.  If we write the Fierz relations
in the form
\begin{equation}
\label{defK}
\bm e(1234) = \bm K^{(abcd)} \bm e(abcd) \,,
\end{equation}
then the matrices $\bm K$ are summarized in Table~\ref{t:scalar}.
\begin{table}
\caption{Fierz matrices for all scalar combinations.  The initial
order of the spinors is taken to be (1234), as expressed in the first
entry. We are omitting those final orderings that, by using the
rule $e_I(abcd) = e_I(cdab)$, reduce to the ones already included.
\label{t:scalar}}
$$
\begin{array}{ll}
\hline
\multicolumn{1}{c}{\mbox{Final order}} &
\multicolumn{1}{l}{\bm K} \\
\hline 
(1234) & \umat \\ 
(1432) & \bm F \\ 
(2^c 1^c 34) & \bm S \\ 
(124^c 3^c) & \bm S \\ 
(13^c 2^c 4) & \bm {SFS} \\ 
(13^c 4^c 2) & \bm {SF} \\ 
(142^c 3^c) & \bm {FS} \\ 
(2^c 1^c 4^c 3^c ) & \bm {SS} = \umat \\
(31^c 2^c 4) & \bm {SF} \\
(31^c 4^c 2) & \bm {SFS} \\
(4^c 1^c 2^c 3^c) & \bm {F} \\
(4^c 1^c 32) & \bm {FS} \\
\hline
\end{array}
$$
\end{table}
The matrices $\bm F$ and $\bm S$ have already been given explicitly in
\Eqs{F}{S}, respectively.  For a ready reference, we give the explicit
forms of their combinations that occur in this list.
\begin{eqnarray}
\label{FS}
\bm {FS} &=& \frac14 \left(\begin{array}{rrrrr}
 -1 &  1 &  \frac12 &  1 & -1 \\
 -4 & -2 & 0       &  2 &  4 \\
-12 &  0 &-2      &  0 &-12 \\
  4 & -2 & 0       &  2 &-4 \\
 -1 & -1 &  \frac12 & -1 &-1
\end{array}\right) \,,\\  
\label{SF}
\bm {SF} &=& \frac14 \left(\begin{array}{rrrrr}
-1  & -1 & -\frac12 &  1 & -1 \\
 4  & -2 & 0        & -2 & -4 \\
 12 &  0 & -2       &  0 & 12 \\
 4  &  2 & 0        &  2 & -4 \\
-1  &  1 & -\frac12 & -1 & -1
\end{array}\right)\,,\\  
\label{SFS}
\bm {SFS} &=& \frac14 \left(\begin{array}{rrrrr}
 1  & -1 &-\frac12 & -1 &  1 \\
-4  & -2 & 0       &  2 &  4 \\
-12 &  0 & -2      &  0 & -12 \\
-4  &  2 & 0       & -2 &  4 \\
 1  &  1 &-\frac12 &  1 &  1
\end{array}\right) \,.
\end{eqnarray}
%

\section{Fierz identities for pseudoscalar combinations}
In Section\ \ref{s:scalar}, we have dealt with scalar quadrilinears
only.  The Fierz transformations are most useful for weak interaction
calculations \cite{weak}, where parity violating pseudoscalar
quadrilinears appear as often as the scalar ones.  Therefore the
transformation matrices for such combinations are also useful.  We
derive them in this section.

We want to consider the quadrilinears of the form indicated 
in \Eqs{genquad1}{genquad2}, and we thus define
\begin{eqnarray}
e'_I (1234) = n_I n_{\tilde I} \, e^r_I(12)e_{\tilde Ir}(34) \,,
\end{eqnarray}
where $\tilde I$ denotes the parity transform of the index $I$, i.e.,
if $I=S,V,A,P$, then $\tilde I = P,A,V,S$ respectively.  In addition,
the matrices $\Gamma^r_{\tilde T}$ have been defined in \Eq{defTdual},
while $n_{\tilde T}\equiv n_T$.  {From} the definition, we can write
down the results for some interchanges immediately.  Denoting by
$(abcd)$ any ordering of the indices, we have, for example,
\begin{eqnarray}
e'_S (abcd) & = & e'_P(cdab) \,, \nonumber\\*
e'_V (abcd) & = & e'_A(cdab) \,, \nonumber\\*
e'_T(abcd) & = & e'_T(cdab) \,.
\label{cdab}
\end{eqnarray}
In order to summarize them in a convenient form, we introduce
the notation
\begin{equation}
\ddiag{d_1,d_2,d_3,d_4,d_5}  \equiv
\left(\begin{array}{rrrrr}
 0 &  0 & 0 & 0 & d_1 \\
 0 &  0 & 0 & d_2 & 0 \\
 0 &  0 & d_3 & 0 & 0 \\
 0 &  d_4 & 0 & 0 & 0 \\
 d_5 &  0 & 0 & 0 & 0 \\
\end{array}\right) \,.
\end{equation}
We can then write
\begin{eqnarray}
\label{defX}
\bm e' (abcd) = {\bm Xe}' (cdab) \,,
\end{eqnarray}
where
\begin{equation}
\label{X}
\bm X = \ddiag{1,1,1,1,1}  \,.
\end{equation}

The other interchanges can be worked out as follows. From the
definitions in \Eqs{defGam}{defTdual}, and remembering \Eq{Tdual=TP},
we find
\begin{eqnarray}
\Gamma^r_A & = & i\Gamma^r_V \Gamma_P\,, \nonumber\\
\Gamma^r_{\tilde T} & = & \Gamma^r_T \Gamma_P \,.
\end{eqnarray}
Then, using also the fact that $\Gamma_P^2 = \umat$, we can write
\begin{eqnarray}
e'_I (1234) = e_I (1234') \,,
\label{4'quads}
\end{eqnarray}
where the notation $4'$ indicates that the spinor that appears in the
fourth position is
\begin{eqnarray}
w'_4 \equiv \Gamma_P w_4 \,.
\label{w'4}
\end{eqnarray}
It is clear that, for any reordering in which the last spinor is not
moved, the same manipulations of Section\ \ref{s:scalar} lead to the
same the Fierz transformation. In other words, if we denote by
$(abc4)$ any such reordering of $(1234)$, \Eqs{defK}{4'quads} imply
that
\begin{equation}
\label{defK4'}
\bm{e}^\prime(1234) = \bm{K}^{(abc4)}\bm{e}^\prime(abc4)\,,
\end{equation}
with the same matrices $\bm K$ tabulated in Table~\ref{t:scalar}.

\begin{table}
\caption{Fierz matrices for all pseudoscalar combinations.  The initial
order of spinors is taken to be (1234).\label{t:pseudoscalar}} 
$$
\begin{array}{ll|||ll}
\hline
\multicolumn{1}{c}{\mbox{Final order}} &
\multicolumn{1}{c|||}{\mbox{Fierz matrix}} &
\multicolumn{1}{c}{\mbox{Final order}} &
\multicolumn{1}{c}{\mbox{Fierz matrix}} 
\\
\hline  \vphantom{\frac {A^{2^3}}2}
(1234) & \umat & (3412) & \bm {X} \\ 
(3214) & \bm F & (1432) & \bm {FX} \\ 
(2^c 1^c 34) & \bm S & (342^c 1^c) & \bm {SX} \\ 
(2^c 3^c 14) & \bm {FS} & (142^c 3^c) & \bm {FSX} \\ 
(31^c 2^c 4) &\bm {SF} & (2^c 431^c) & \bm {SFX} \\
(13^c 2^c 4) &\bm {SFS} & (2^c 413^c) & \bm {SFSX} \\ 
\hline
(124^c 3^c) & \bm {XSX} & (4^c 3^c 12) & \bm {XS} \\ 
(13^c 4^c 2) & \bm {XSF} & (4^c 213^c) & \bm {XSFX} \\ 
(2^c 1^c 4^c 3^c ) & \bm {SXSX} & (4^c 3^c 2^c 1^c) & \bm{SXS} \\
(324^c 1^c) & \bm {XFSX} & (4^c 1^c 32) & \bm {XFS} \\
(31^c 4^c 2) & \bm {XSFS} & (4^c 231^c) & \bm {SFXS} \\
(2^c 3^c 4^c 1^c) & \bm {SXSF} & (4^c 1^c 2^c 3^c) & \bm {FSXS} \\ 
\hline
\end{array}
$$
\end{table}
With the help of \Eqss{defK}{defX}{defK4'} we can find all the
remaining transformations in terms of the matrices $\bm F$, $\bm S$
and $\bm X$, and their products.  For example, the analog of
\Eq{fierzstandardshort} can be obtained through the following steps,
\begin{eqnarray}
\bm e' (1234) = \bm {Xe}' (3412) = \bm {XF e'} (1432) \,,
\end{eqnarray}
and the analog of \Eq{S34} would be obtained from
\begin{eqnarray}
\bm e' (1234) = \bm {Xe}' (3412)  = \bm {XSe}' (4^c3^c12) 
= \bm {XSXe}' (124^c3^c) \,.
\end{eqnarray}
In analogy with \Eq{defK}, we summarize these transformations by writing
\begin{equation}
\label{defK'}
\bm e'(1234) = \bm K^{\prime\,(abcd)} \bm e'(abcd) \,,
\end{equation}
where the complete list of the matrices $\bm K'$ is tabulated in
Table~\ref{t:pseudoscalar}. Apart from those entries in which only one
of the matrices $\bm F$, $\bm S$ or $\bm X$ appear, the product
matrices can be split in two groups according to whether or not the
product contains the matrix $\bm X$. The explicit expressions of the
products that do not contain any factor of $X$ have already been given
in \Eqss{FS}{SF}{SFS}.  Those that contain one or more factors of $\bm
X$ are given by
\begin{eqnarray}
\label{XSX} 
\bm {XSX} &=& \diag{-1,-1,+1,+1,-1} \,, \\
\label{SXS} 
\bm {SXS} &=& \ddiag{+1,-1,+1,-1,+1} \,, \\
\label{XFS}
\bm{XFS} &=& \frac14 \left( \begin{array}{rrrrr}
 -1 & -1 & \frac12 & -1 &  -1\\  
  4 &  -2 & 0       & 2 &  -4\\  
-12 &  0 & -2      &  0 & -12\\  
 -4 & -2 & 0       &  2 &   4\\  
 -1 &  1 & \frac12 &  1 &  -1
			    \end{array} \right) \,, \\
\label{XSF}
\bm {XSF} &=& \frac14 \left(\begin{array}{rrrrr}
-1  &  1 & -\frac12 & -1 & -1 \\
 4  &  2 & 0        &  2 & -4 \\
 12 &  0 & -2       &  0 & 12 \\
 4  & -2 & 0        & -2 & -4 \\
-1  & -1 & -\frac12 &  1 & -1 
\end{array}\right)\,,\\  
\bm{FSXS} &=& \frac14 \left( \begin{array}{rrrrr}
  1 &  1 & \frac12 & -1 &  1\\  
 -4 &  2 & 0       &  2 &  4\\  
 12 &  0 & -2      &  0 & 12\\  
  4 &  2 & 0       &  2 & -4\\  
  1 & -1 & \frac12 &  1 &  1
			     \end{array} \right)\,, \\
\bm{SXSF} &=& \frac14 \left( \begin{array}{rrrrr}
  1 & -1 & \frac12 &  1 &  1\\  
  4 &  2 & 0       &  2 & -4\\  
 12 &  0 & -2      &  0 & 12\\  
 -4 &  2 & 0       &  2 &  4\\  
  1 &  1 & \frac12 & -1 &  1
			     \end{array} \right)\,, \\
\bm{SFXS} &=& \frac14 \left( \begin{array}{rrrrr}
  1 &  1 & -\frac12 &  1 &   1\\  
  4 & -2 & 0        &  2 &  -4\\  
-12 &  0 & -2       &  0 & -12\\  
  4 &  2 & 0        & -2 &  -4\\  
  1 & -1 & -\frac12 & -1 &   1
			     \end{array} \right) \,.
\end{eqnarray}
The remaining ones are of the form of one of the matrices already
given, multiplied by the matrix $\bm X$ to the left or to the right.
Multiplying a matrix by $\bm X$ to the right is equivalent to reading
its columns backwards and, similarly, multiplying a matrix by $\bm X$
to the left amounts to reading its rows backwards. In either case, the
result can be easily read off the explicit expressions given above and
in Section\ \ref{s:scalar}.  In addition, it is helpful to note that
the Fierz matrices that appear in Table~\ref{t:pseudoscalar} may be
rewritten in alternative forms, using the relations
\begin{eqnarray}
\bm{FX} &=& \bm{XF} \,, \nonumber\\ 
\bm{SXSX} &=& \bm{XSXS} \,, \nonumber\\ 
\bm{XFSXS} &=& \bm{SXSF} \,, \nonumber\\ 
\bm{XSFSX} &=& \bm{SFXS} \,,
\end{eqnarray}
and similar ones.
They can be derived in much the same way as \Eq{SFS=FSF}.

\section{Examples and Summary}
The manipulations involved in the use of the Fierz transformations are
useful in practical calculations, but the question of when and how to
use them can only be answered in the context of the specific situation
at hand. Thus, while it is not possible to give a single answer to
those questions, we illustrate, with some examples, how we can judge
if it is prudent to use them, and how.

\subsection{Invariants}
A well known result is that the combinations of quadrilinears
\begin{eqnarray}
\label{Qpm}
Q_{\eta}(1234) \equiv 
\Big[ \bar w_1 \gamma^\mu (1 +  \eta\gamma_5) w_2 \Big] 
\Big[ \bar w_3 \gamma_\mu (1 + \eta\gamma_5) w_4 \Big] \,,
\end{eqnarray}
where $\eta = \pm 1$,
are form invariant under the standard Fierz transformation.
In our notation, they can be expressed in the form
\begin{eqnarray}
Q_{\eta}(1234) = e_V(1234) + e_A(1234) + 
\eta\left(e'_V(1234) + e'_A(1234)\right)\,,
\end{eqnarray}
and the form invariance statement is the property that
\begin{equation}
\label{invQpm}
Q_{\eta}(1234) = \null - Q_{\eta}(1432) \,.
\end{equation}
(If we were dealing with field operators instead of numerical spinors,
the overall minus sign in \Eq{invQpm} would have been absent.)  The
obvious question is whether there are other combinations that share
this property, and not just under the standard transformation, but
also under the generalized Fierz transformations that we have
considered. The answer to this question is affirmative.  We illustrate
this below with a few examples, in a way that shows how we can make
more general statements, in a systematic fashion.

Let us denote an arbitrary combination of the scalar quadrilinears by
\begin{eqnarray}
Q_{\{a\}} \equiv \sum_I a_I e_I (1234)\,,
\label{Qa}
\end{eqnarray}
where the $a_I$ are numerical coefficients.  Using \Eq{defK}, it can be
expressed with a different spinor ordering as
\begin{eqnarray}
Q_{\{a\}} = \sum_{I,J} a_I K^{(abcd)}_{IJ} e_J(abcd) \,.
\end{eqnarray}
This will be of the same form as in \Eq{Qa} if
\begin{eqnarray}
\sum_I a_I K^{(abcd)}_{IJ} & = & \lambda a_J\,,
\end{eqnarray}
where $\lambda$ is a numerical factor. That is, the row matrix
composed of the elements $a_I$ is a left eigenvector of the matrix
$\bm K^{(abcd)}$ with the eigenvalue $\lambda$ or, equivalently, the
column matrix composed of the elements $a_I$ is a right eigenvector of
the matrix $[\bm K^{(abcd)}]^\top$ with the eigenvalue
$\lambda$. Since the transformation matrices are not symmetric
(strictly speaking, they are not normal matrices), the distinction
between right and left eigenvectors is important.

Similarly, we define an arbitrary pseudoscalar combination by
\begin{eqnarray}
Q'_{\{a'\}} \equiv \sum_I a'_I e'_I (1234)\,.
\label{Q'a}
\end{eqnarray}
Then using \Eq{defK'},
\begin{eqnarray}
Q'_{\{a'\}} = \sum_{I,J} a'_I K^{\prime\,(abcd)}_{IJ} e'_J(abcd) \,,
\end{eqnarray}
so that $Q'_{\{a'\}}$ will retain its form if
\begin{eqnarray}
\sum_I a'_I K^{\prime\,(abcd)}_{IJ} & = & \lambda' a'_J\,,
\end{eqnarray}
which indicates that the column matrix composed of the elements $a'_I$
is a right eigenvector of the matrix $[\bm K^{\prime\,(abcd)}]^\top$
with the eigenvalue $\lambda'$.  Notice that, for those invariant
combinations $Q_{\{a\}}$ and $Q'_{\{a'\}}$ such that $\lambda =
\lambda'$, then any linear combination
\begin{equation}
\label{QQ'}
\alpha Q_{\{a\}} + \beta Q'_{\{a'\}}
\end{equation}
is form invariant as well.

As an example, we consider the final ordering $(abcd) = (1432)$, for
which the transformation matrices $\bm K$ and $\bm K'$ are equal to
$\bm F$. To find the eigenvectors of $\bm F^\top$, we observe that
$\bm F^2 = 1$, which implies that the eigenvalues are $\pm
1$. Further, since $\mbox{Tr}\,\bm F = -1$, it follows that the
eigenvalue $+1 (-1)$ is doubly (triply) degenerate. A particular
choice of the eigenvectors is listed below, along with the
corresponding eigenvalues:
\begin{eqnarray}
\begin{array}{cc} \hline 
\mbox{Eigenvector} & \mbox{Eigenvalue} \\ \hline 
(0,1,0,1,0) & -1 \\
(2,0,-1,0,2) & -1 \\
(-2,1,0,-1,2) & -1 \\
(1,0,\frac16,0,1) & +1 \\
(2,1,0,-1,-2) & +1 \\ \hline
\end{array}
\end{eqnarray}
Thus, for example, the first line indicates that the combination
\begin{equation}
e_V + e_A
\end{equation}
is invariant, and so is, separately, the combination
\begin{equation}
e'_V + e'_A \,.
\end{equation}
By \Eq{QQ'}, any linear combination
\begin{equation}
\alpha(e_V + e_A) + \beta(e'_V + e'_A)
\end{equation}
is also invariant, which includes the chiral combinations
given in \Eq{Qpm} as special cases.

The others show other invariants.  For example, the last line of the
table shows that
\begin{eqnarray}
&& 2e_S + e_V - e_A -2e_P\,,\nonumber\\
&& 2e'_S + e'_V - e'_A -2e'_P \,,
\end{eqnarray}
and any linear combination of them are invariant as well.

We can apply the same method to the other transformation matrices.
For example, consider the final ordering $(abcd) = (1423)$, for which
$\bm K = \bm{FS}$ while $\bm K^\prime = \bm{FSX}$.  Since this
permutation, performed thrice, brings back the original ordering, we
must have $(\bm{FS})^3 = 1$, which can also be checked explicitly from
the matrix given in \Eq{FS}.  Consequently, the eigenvalues are the
cube roots of 1; i.e., the only (non-degenerate) real eigenvalue is 1,
with the corresponding eigenvector $(1,0,0,1,-1)$.  Thus, in
particular, $(e_S + e_A - e_P)$ is invariant under the interchange
being considered.  For the pseudoscalars ($[\bm{FSX}]^T$), also there
is only one real eigenvalue (+1) with the eigenvector
$(1,1,0,0,-1)$. Therefore $(e'_S + e'_V - e'_P)$ is invariant, and by
\Eq{QQ'} so is any combination of the form
\begin{equation}
\alpha (e_S + e_A - e_P) + \beta (e'_S + e'_V - e'_P) \,.
\end{equation}
\subsection{Chiral quadrilinears}
By chiral quadrilinears we mean those in which only one chiral
component of each spinor appears.  Consider, for example,
\begin{equation}
Q_{RL} \equiv \Big[ \bar u_1 R v_2 \Big]
\Big[ \bar v_3 L u_4 \Big]\,.
\end{equation}
where, as usual,
\begin{eqnarray}
R & = & \frac{1}{2}(1 + \gamma_5)\, \nonumber\\
L & = & \frac{1}{2}(1 - \gamma_5) \,.
\end{eqnarray}
This can be brought to a form in which only $u$-type spinors appear by
combining the transformations given in Tables \ref{t:scalar} and
\ref{t:pseudoscalar}, but in this case it is more economical to
proceed in the following way.  Decomposing each spinor $w$ in terms of
its chiral components,
\begin{equation}
w = w_L + w_R \,,
\end{equation}
with $w_L = Lw$ and $w_R = Rw$, we can write
\begin{eqnarray}
Q_{RL} & = & \Big[ \bar u_{L1} v_{R2} \Big]
\Big[\bar v_{R3} u_{L4} \Big]\nonumber\\
& \equiv & e_S(1234) \,,
\end{eqnarray}
where the index $1$ is associated with th spinor $u_{L1}$, the index 2
with $v_{R2}$, and so on.  This can be rewritten with the spinors in
the order $(1423)$ by means of the matrix $\bm{FS}$ from Table
\ref{t:scalar}; i.e.,
\begin{eqnarray}
e_S(1234) & = & \frac{1}{4}\left[
-e_S(142^c 3^c) + e_V(142^c 3^c) +
\frac{1}{2}e_T(142^c 3^c)\right.\nonumber\\
&&\mbox{}\left.\vphantom{\frac{1}{2}} + e_A(142^c 3^c)
- e_P(142^c 3^c)\right] \,.
\end{eqnarray}
The definition \Eq{what} implies that, for the chiral components,
\begin{eqnarray}
(u_{L,R})^c & = & v_{R,L} \,,\nonumber\\
(v_{L,R})^c & = & u_{R,L} \,,
\end{eqnarray}
and using the fact that
\begin{eqnarray}
\bar u_L\Gamma_{S,T,P}\,u_L & = & 0\nonumber\\
\bar u_L\Gamma_A\,u_L & = & -\bar u_L\Gamma_V\, u_L \,,
\end{eqnarray}
we finally arrive at
\begin{equation}
\Big[ \bar u_1 R v_2 \Big]\Big[ \bar v_3 L u_4 \Big] = \frac{1}{2}
\Big[ \bar u_1 \gamma^\mu L u_4 \Big]\Big[\bar u_2\gamma_\mu L
u_3\Big] \,. 
\end{equation}

In practice however, the amplitudes generally consist of a superposition
of terms so that, instead of the single term $Q_{RL}$, the relevant
quantity to consider would likely be of the form
\begin{equation}
\Big[\bar u_1 (a + b\gamma_5) v_2 \Big]
\Big[\bar v_3 (c + d\gamma_5) u_4 \Big]\,.
\end{equation}
This could be manipulated by using the chiral decomposition of the
spinors, which would generate several chiral quadrilinears ($Q_{RL},
Q_{LL}$ and so on), and then proceeding as above with each of them.
While such manipulations can of course be readily mastered, the
transformations given in Tables \ref{t:scalar} and
\ref{t:pseudoscalar} provide a ready-to-use recipe for cases such as
this, and more general ones.

\end{document}